**FINAL DRAFT – Submitted to eContact! (https://econtact.ca)**

**Special issue: 21.1 — Take Back the Stage: Live coding, live audiovisual, laptop orchestra…**

# Virtual Agents in Live Coding: A Short Review

## Past, Present and Future Directions

by Anna Xambó

A short review of past, present and future directions of virtual agents in live coding.

## Introduction

Machine learning is a trending topic in science, engineering and the arts (Alpadyn 2016). Bringing machine learning into the real-time domain is a well-known challenge due to the expected time for computations (Collins 2008, Xambó et al. 2019). Artificial intelligence (AI) and improvisation have been investigated extensively, including music improvisation. However, with a few notable exceptions discussed here, AI and the improvisational practice of live coding has been little explored. Live coding is an exemplary scenario to investigate the potential of machine learning in performance because it brings a live account of programming applied to music-making, described with the property of *liveness* (Tanimoto 2013).

This article looks into the present and future directions that research and practice on virtual agents (VAs) in live coding might take. As discussed at the beginnings of live coding in 2003, beyond the promises and potential of a *"new form of expression in computer music"* (Collins et al. 2003, 321), the drawbacks and dangers of live coding were also highlighted, including the amount of time required to generate audible code, the dependency on inspiration which is not always present when improvising, and the risk of dealing with code errors due to its liveness. Collaborative music live coding (CMLC) typically involves a group of at least two networked live coders, who can be either co-located in the same space, distributed in different spaces, or both (Barbosa 2006). CMLC is a promising approach to music performance and computer science education because it can promote peer-learning in the latter, and an egalitarian approach to collaborative improvisation in the former (Xambó et al. 2017). Sometimes it is unpractical to collaborate with other humans, yet the social need remains.

This article contributes with a short review of different perspectives of using VAs in the practice of live coding from past and present, focusing on the last few years and pointing to future directions. Different database sources have been used (DBLP, Google Scholar, ICLC Proceedings, JSTOR Search, Scopus, Zenodo) using different keywords *("live coding + machine learning"*, *"live coding + agent"*, *"live coding + virtual agent"*), especially looking at publications from the latest decade (2010–2020). We have excluded references that explore immersive experiences (e.g. virtual worlds, audiovisual experiences) as it is out of the scope of this research. Our analytical approach is based on thematic analysis (Braun and Clarke 2006) using the NVivo software [[1. https://www.qsrinternational.com/nvivo-qualitative-data-analysis-software/home]] to identify the main topics discussed here. The total number of analyzed references is 30 items (2003–2020), excluding from this count those references more generic and contextual. The references have been grouped into two main categories: (1) *conceptual foundations*; and (2) *practical examples*. Figure 1 illustrates the most frequent specialized words, which orbit around vocabulary from computer music, computing, and machine learning, among others.

This article is a follow-up to our previous research on exploring the potential role of a virtual agent to counterbalance the limitations of human live coding (Xambó et al. 2017). In turn, it provides a critical context to position our EPSRC HDI Network Plus funded project “MIRLCAuto: A Virtual Agent for Music Information Retrieval in Live Coding” [[2. https://mirlca.dmu.ac.uk]]. Contributing to the music performance domain can more broadly inform the emerging fields of machine learning and computational creativity applied to real-time contexts (e.g. creative computing, music education, music production, performing arts) and help to move the field forward.

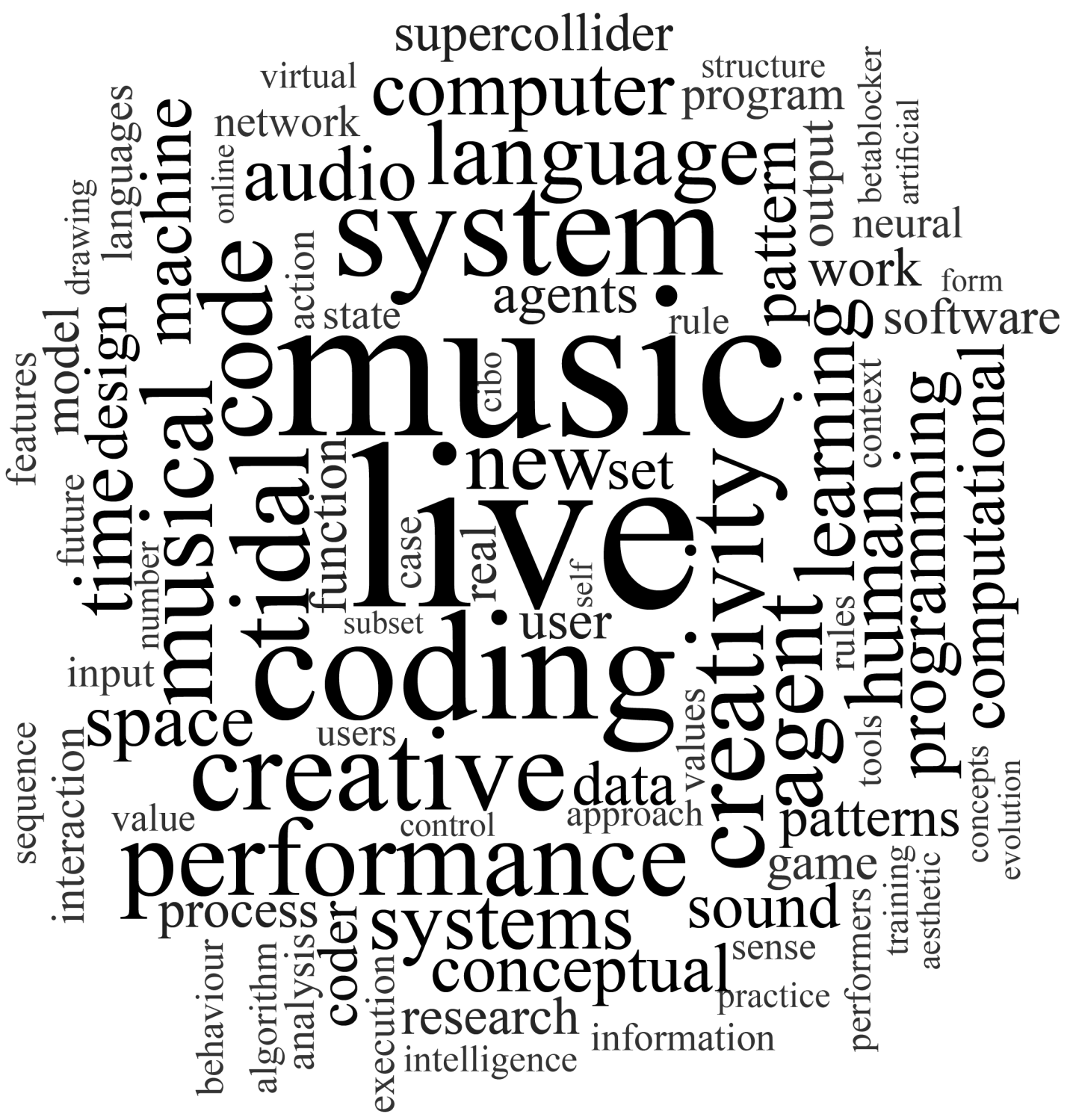

Figure 1: Word cloud that shows the 100 most frequent specialized words used in the analyzed references related to virtual agents in live coding. Generated with NVivo.

## Conceptual Foundations

### Broader Context

Several areas inform the field of VAs in live coding. Figure 2 shows the main keywords that are present in the selected readings. These include interdisciplinary fields (e.g. affective computing, genetic programming, human-computer interaction, music technology), fields of AI and creativity (e.g. computational creativity), fields of AI and computer music (algorithmic composition, computer-aided composition, laptop orchestra, machine musicianship), approaches to collaboration and AI (e.g. co-creation, multi-agent systems, participatory sense-making), and approaches to interactive music systems (e.g. musical interface design), among others. This variety illustrates the interdisciplinary character of the field.

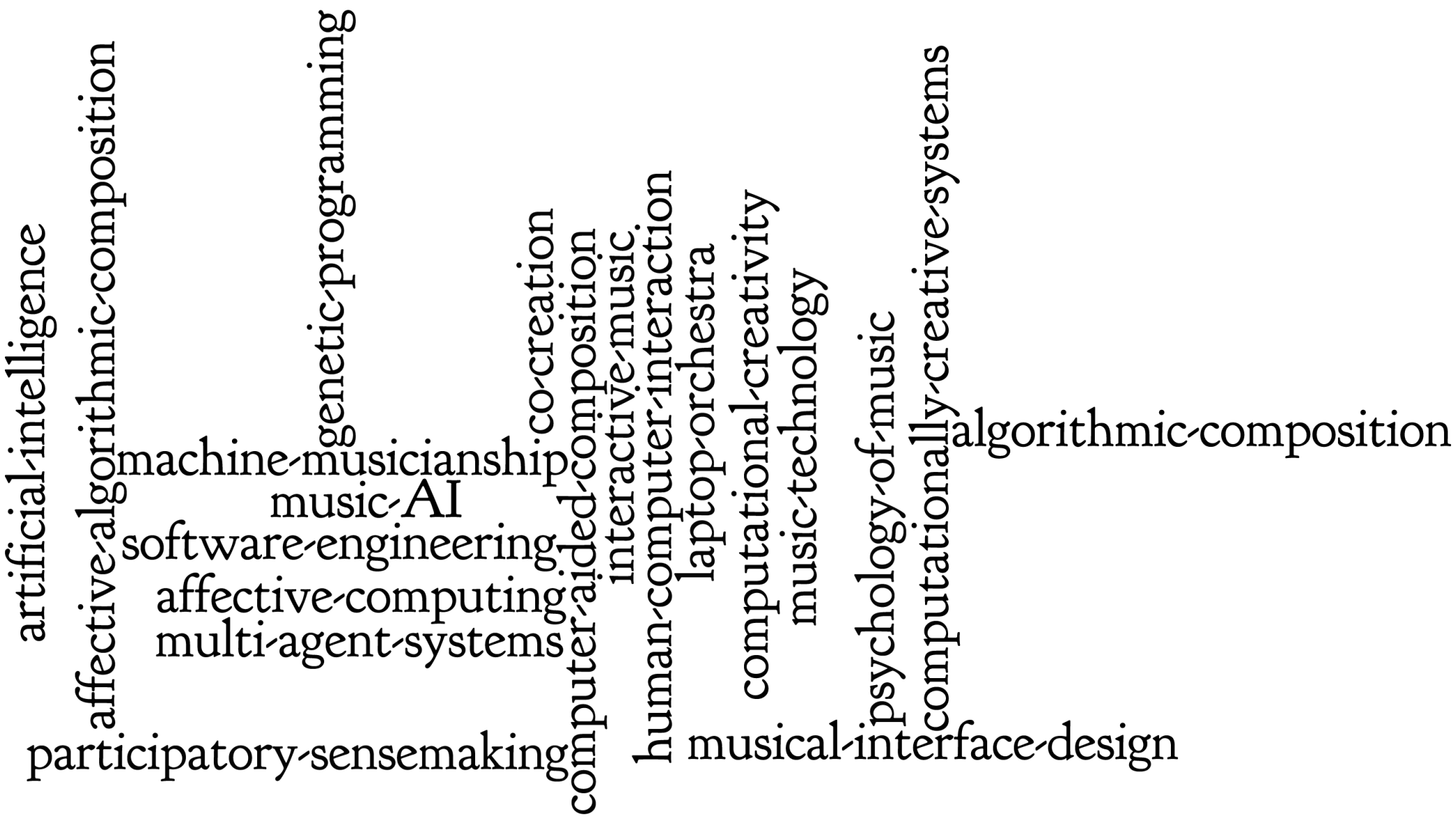

**Figure 2: Word cloud that shows related disciplines and perspectives that inform virtual agents in live coding. Generated with Wordle.**

## Virtuality, Agency, and Virtual Agency

In this section, we present the working definitions for the terms “virtual”, “agent” and “virtual agent”.

The noun virtual comes from the Latin *virtus* ‘virtue’, meaning courage, bravery, strength, power, worth, manliness, virtue, character or excellence.

> Virtuality is the property of a computer system with the potential for enabling a virtual system (operating inside the computer) to become a real system by encouraging the real world to behave according to the template dictated by the virtual system. In philosophical terms, the property of virtuality is a system’s potential evolution from being descriptive to being prescriptive. (Turoff 1997, 38)

From this perspective, virtuality refers to how computers represent reality. The term *virtual* has been used a long time before the appearance of computers and has multiple connotations, including the related term ‘virtue’ as a personal quality, and the ‘liminal’ property in virtual spaces which refers to temporary zones (Shields 2003, 1–17).

The noun agent comes from the Latin agent- ‘doing’, from *agere*, meaning to put in motion, move, lead, drive, act, tend or conduct. A commonly accepted definition of the term *agent* is “one who acts, or who can act” (Franklin and Graesser 1996, 25), which includes humans and most animals, but also computer-based agents. More precisely, an agent is defined as *“anything that can be viewed as perceiving its environment through sensors and acting upon that environment through actuators.”* (Russell and Norvig 2016, 35) A more specific working definition brings the notion of being capable of acting independently:

> An autonomous agent is a system situated within and a part of an environment that senses that environment and acts on it, over time, in pursuit of its own agenda and so as to effect what it senses in the future. (Franklin and Graesser 1996, 25)

According to Russel and Norvig, *“for each possible percept sequence [the complete history of everything the agent has ever perceived], a rational agent should select an action that is expected to maximize its performance measure, given the evidence provided by the percept sequence and whatever built-in* *knowledge the agent has"* (Russell and Norvig 2016, 37), hence *"a rational agent should be autonomous"* (Russell and Norvig 2016, 39).

A commonality of these definitions of an agent is to cover from the most complex systems on the one end, e.g. humans with sophisticated senses and a range of possible actions to choose from, to the simplest systems on the other end, e.g. thermostats, which are just reactive because they simply react to the temperature of the environment. Russel and Norvig term the simplest type of agent as a *simple reflex agent*, which only considers what it senses at present and *"have the admirable property of being simple, but they turn out to be of limited intelligence"* (Russel and Norvig, 2016, 49). Accordingly, it is worth noting that a program is not necessarily a computer agent if it does not continuously sense the environment and it does not affect what it senses in a later period. As described by Russell and Norvig:

> Computer agents are expected to do more [than just act]: operate autonomously, perceive their environment, persist over a prolonged time period, adapt to change, and create and pursue goals. (Russell and Norvig 2016, 4)

A software rational agent is often referred to as a virtual agent, which is the term that we use in this article. We are cautious about using the term 'intelligent agent' as it can raise unrealistic expectations (Nwana 1996), although the state-of-the-art is progressing rapidly. From the above, we assume that VAs are autonomous, and they can span from simple to complex types.

Three primary non-exclusive binary attributes have been identified as preferred criteria in software agents: (1) *autonomy*, which refers to the capacity of agents to operate on their own; (2) *cooperation*, which refers to the ability to interact with other agents including humans; and (3) *learning*, which refers to the ability to learn in the process of reacting or interacting with the environment (Nwana 1996). Russell and Norvig (2016) distinguish three aspects related to what can constitute a rational agent: (1) *omniscience*, which refers to the unrealistic and unwanted ability to know the outcome of future actions (not to confuse with rationality and the ability to gather or explore information); (2) *learning*, which refers to the ability to learn from what it is perceived; and (3) *autonomy*, which refers to the ability to learn and modify prior knowledge. Accordingly, one of the advantages of a learning agent is that *"it allows the agent to operate in initially unknown environments and to become more competent than its initial knowledge alone might allow"* (Russell and Norvig 2016, 55). This is possible via the additional learning element, which is in charge of improving the agent's performance through constantly modifying its behaviour according to feedback and learning goals. The learning component builds on the performance element, which is in charge of deciding actions and is always present in the agents discussed here.

## Machine Musicianship, Musical Metacreation, and Musical Cyborgs

Other relevant terms related to VAs in computer music include machine musicianship, musical metacreation and musical cyborg.

*Machine musicianship* applies artificial intelligence concepts and techniques to computer music systems, where the systems can learn and evolve (Rowe 2001). Some notable examples are: (1) Voyager, which is an interactive musical system with the ability to improvise with a human improviser by combining responses and independent behaviours (Lewis 2000); (2) the Continuator, which is a musical instrument that learns and plays interactively with the performer's style (Pachet 2003); and (3) Shimon, which is a robotic musician that can improvise with humans (Hoffman, Guy and Weinberg 2010). There has been an interest to study the application of machine musicianship in live coding (Subramanian et al. 2012, Wilson et al. 2020). In our research, we are particularly interested in exploring a virtual agent companion that learns from human live coders, using machine learning algorithms, and that goes beyond the approach of following live coder actions (also known as the call-response strategy). To embody the humanoid metaphor, we envision that the virtual agent should be able to act both as a live coder and a chatting peer (Xambó et al. 2017).

A specialized exploration of agents in computer music, in terms of computational simulations of musical creativity, can be found at the annual International Workshop on Musical Metacreation Workshop [[3. http://musicalmetacreation.org]]. Connected with the creative practice field of artificial life art or metacreation (Whitelaw 2004), the field of musical metacreation investigates generative tools and theories applied to music creation and includes in its research collaboration between humans and creative VAs (Pasquier et al. 2016). In live coding, CMLC can be seen as a conversation between at least two people, which can include VAs in addition to human agents (HAs). Beyond VA-HA collaboration, we envision VA-VA collaboration (Xambó et al. 2017). Accordingly, several research questions emerge: Can multiple agents collaborate among themselves? How feasible is it? What would be the computational cost? To what extent there should be supervision and how often? Collaboration between live-coding VAs can be seen as a particular case of multi-agent systems for music composition and performance, which has been widely researched (Miranda 2011).

Donna Haraway has accurately described the term *cyb/ernetic org-anism* or *cyborg* as: *"a cybernetic organism, a hybrid of machine and organism, a creature of social reality as well as a creature of fiction."* (Haraway 1991, 150). This creature of social and fictional realities has profoundly influenced science, engineering and the arts, including music and live coding. The *musical cyborg* has been defined as *"a figure which combines human creativity and digital algorithms to create sounds and moments that would not otherwise be possible through human production alone."* (Witz 2020, 33). This close relationship between the live coder and the algorithm is further described: *"the live coded cyborg actively shapes the algorithms they use to optimize their output on an individual level."* (Witz 2020, 35). Accordingly, here we use the term *musical cyborg* as a metaphor that refers to the cooperation between a human live coder and a virtual agent, particularly when it is a one-to-one relationship.

## Theoretical Frameworks

In this section, we revise a number of theoretical frameworks that can help design and evaluate systems of VAs for live coding.

A general definition of creativity is *"the ability to generate novel, and valuable, ideas"* (Boden 2009, 24). However, assessing whether a computer is creative is in itself a philosophical inquiry (Boden 2009). In computational creativity, we find some theoretical frameworks designed for understanding computational creative systems e.g. the Standardised Procedure for Evaluating Creative Systems (SPECs) (Jourdanous 2012) and the Creative Systems Framework (CSF) (Wiggins 2006), among others. In particular, CSF has been discussed in the context of live coding (McLean and Wiggins 2010, Wiggins and Forth 2018). Transformational creativity is highlighted as an important element for an agent to be considered creative, *"in which an agent modifies its own behaviour by reflective reasoning."* (Wiggins and Forth 2018, 274).

Tanimoto provides six levels of liveness in the process of programming (Tanimoto 2013). This theoretical framework looks into the relationship between the programmer's actions and the computer's responses. Beyond the previously discussed four levels of *"informative"*, *"informative and significant"*, *"informative, significant and responsive*" and *"informative, significant, responsive and live"*, two new levels are proposed. Level 5 is called *"tactically predictive"* and is slightly ahead of the programmer by predicting the next programmer's action (e.g. lexical, semantic, musical) utilizing machine learning techniques. Level 6 is foreseen as *"strategically predictive"*, which indicates more intelligent predictions about the programmer's intentions. Intelligent predictions are linked to agency and liveness, where not only the code but also the tool entail liveness:

> The incorporation of the intelligence required to make such predictions into the system is an incorporation of one kind of agency – the ability to act autonomously. Agency is commonly associated with life and liveness. (One might argue that here, liveness has spread from the coding process to the tool itself.) (Tanimoto 2013, 34)

Other approaches to understanding VAs for live coding borrow concepts from studies on collaboration between humans from cognitive science and HCI, among others. A suitable model that takes into account human collaboration and improvisation is the *enactive paradigm* (Davis et al. 2015). This paradigm incorporates an enactive model of collaborative creativity and co-creation to describe improvised collaborative interactions with feedback from the environment. As part of this paradigm, participatory sense-making is situated as a key component, described as: *"negotiating emergent actions and meaning in concert with the environment and other agents"* (Davis et al. 2015, 114). In the context of a co-creative drawing partner, a set of design recommendations for co-creative agents is presented in alignment with participatory sense-making and open-ended collaborations (Davis et al. 2016). Accordingly, co-regulation of interaction between agents should form part of the process of participatory sense-making. As pointed in the TOPLAP manifesto [[4. https://toplap.org/wiki/ManifestoDraft]], "obscurantism can be dangerous" and screens are typically shown. The live coder's actions and algorithmic thoughts can be shared not only with co-creative agents but also with the audience, who can also become part of this process of participatory sense-making.

## Examples in Practice

Scholars have proposed several agent typologies, notably (Nwana 1996, Russell and Norvig 2016). Our assumptions here are that the VAs are autonomous (ranging from simple to complex agents). In this section, we analyze examples of VAs in live coding based on two dimensions: (1) *social interactivity*, does it cooperate with other agents, either virtual or human? and (2) *learnability*, does it learn, either online or offline? *Shared collective control* in computer music has been described with two categories (Jordà 2005): (1) *multiplicative actions*, the product of individual contributions, or a series of highly interdependent processes; and (2) *summative actions*, the sum of individual contributions, or independent processes, where there is little mutual interaction. In this article, we consider these two categories when describing social interactivity to reflect more precisely the possibilities of collaboration in musical practice. Figure 3 summarizes the examples analyzed in this section according to our two-dimensional representation. In the text, we present the examples in chronological order based on the year of the related publication.

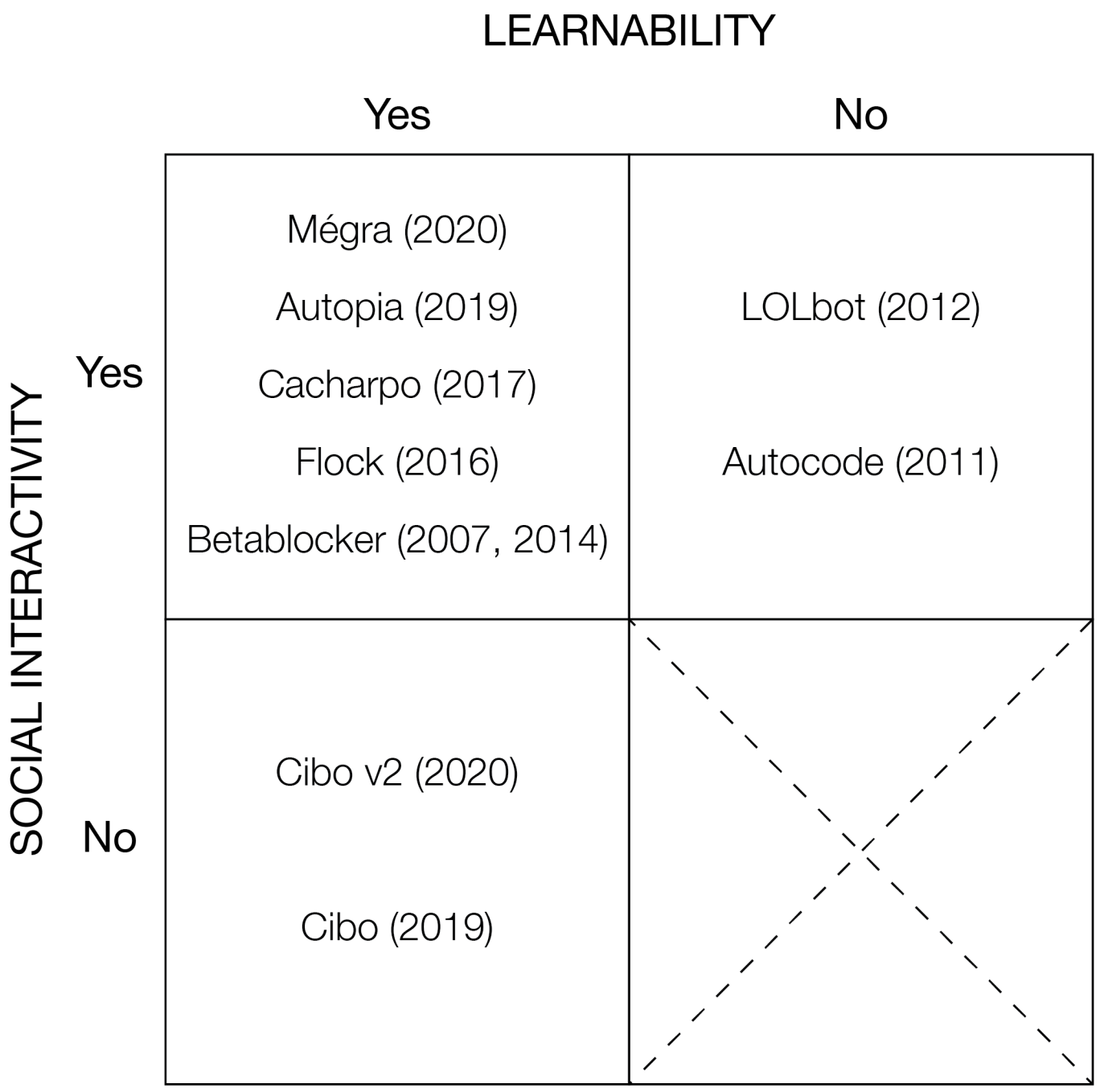

**Figure 3: Matrix of learnability and social interactivity with examples of virtual agents in live coding.**

## Betablocker

Betablocker is a multi-threaded virtual machine designed for low-level live coding that can be used for algorithmic composition, as well as for sonification and visualization. It was originally developed using the gaming engine Fluxus and operated with a gamepad instead of a keyboard and mouse (Griffiths 2007). Later it was ported to the Gameboy DS system (Bovermann and Griffiths 2014). A follow-up version was implemented in SuperCollider (Bovermann and Griffiths 2014). The engine only stops if it is externally halted, but it might evolve to stable repetitive behaviours. The language is highly constrained with no distinction between program and data. With the Fluxus and Gameboy DS versions, it is possible to create several interactive software agents that modify themselves and each other. For example, one program plays sound in a loop, whilst another program overwrites parts of the former while it is playing according to an XOR operation. Betablocker also uses genetic algorithms with a customised fitness function for selecting the next generation of programs, which results in surprising standalone programs. The SuperCollider version results in a more deterministic approach, although this is counterbalanced with randomization functions.

Betablocker is an example of an environment that hosts simple autonomous agents that can interact with other agents (multiplicative actions) as well as the ability to learn using evolutionary algorithms. In the words of the system's authors: *"Betablocker can be viewed as a companion for live coding that one has the opportunity to get to know, collaborate with, and—sometimes—work against."* (Bovermann and Griffiths 2014, 52).

## ixi lang

ixi lang is a live coding language implemented in SuperCollider (Magnusson 2011). The language has been designed for beginners in live coding. The syntax is simple and allows the live coder to manipulate musical patterns in a constrained environment. The interaction metaphor consists of creating agents that have assigned instruments, such as melodic, percussive, sample-based or custom, which can play scores defined by the musical patterns. Autocode [[5. https://youtu.be/jWvzCzR_tus]] is an autonomous and deterministic virtual agent within ixi lang. When invoking the VA, it is possible to define the number of agents or 'musicians' to be created. Then the 'musicians' start to be generated in real time with assigned musical patterns according to a deterministic random behaviour that chooses the type of instrument and the score to be played, among others. Although multiple agents can be created who can play in sync, their processes seem to be independent of each other. Autocode exemplifies a VA live coder who behaves autonomously by creating musical agents with the ability to cooperate (summative actions), but with no ability to learn.

## LOLbot

LOLbot (Subramanian et al. 2012) is a VA implemented in Java that resides in the LOLC environment. LOLC is a text-based environment for collaborative live-coding improvisation. In a performance setting, LOLbot is run on a separate computer and appears as another ensemble member in the interface. The motivation is threefold: to understand the nature of human performance, to bring a new character to the ensemble, and to provide a tool for practising. Time synchronization between performers is kept via a shared clock in the LOLC server. The live coders can create rhythmic patterns based on available sound samples. LOLC has an interface with an instant-messaging feature and a visualization of the created patterns, which can be shared and borrowed (multiplicative actions). LOLbot observes and analyses what the human performers are doing, which is encapsulated as patterns. Then the VA selects suitable patterns to play in the ensemble using the LOLC syntax. LOLbot uses pattern-matching algorithms to identify the preferred choice to play, which is determined by the metric coherence/contrast. This metric is defined, in real time, by the human performers using a slider. The value of the coherence/contrast slider defines the character of the agent between a rhythmic behaviour and a contrasting behaviour. LOLbot is an example of an autonomous agent that can interact with other agents (in this case humans), but with no internal feedback, a mechanism to make improvements.

## Flock

Flock (Knotts 2016) is a system implemented in SuperCollider's JITlib. The system has voting agents who 'listen' to the music made by real human live coders and vote their preferred audio stream according to their musical taste. The audio analysis uses machine listening techniques employing the SCMIR library in SuperCollider. The votes affect the audio level presence of the audio streams in the audio mix. The agents' preferences can change over time depending on the other agents' votes inspired by a bi-partisan political model and flock theory in decentralized networks. Flock exemplifies a system with autonomous agents that can interact with other agents, as well as the ability to constantly modify their behaviour according to the changes in the environment, including other agents' behaviours (multiplicative actions).

## Cacharpo

Cacharpo (Navarro and Ogborn 2020) is a VA capable of live coding that works as a co-performer of a human live coder. The music genre is inspired by the cumbia sonidera from Mexico and with roots from Colombian cumbia. The motivation is to provide a companion that can bring new dynamics into solo or group performances. The agent 'listens' to the audio produced by the live coder using machine listening and music information retrieval techniques implemented in SuperCollider. Artificial Neural Networks (ANNs) developed in Haskell are used in real time to identify the characteristics of the music (e.g. cumbia sonidera roles, instruments, and relevant audio features). The training of the ANNs is performed previously offline using a dataset of sound recordings from SuperCollider performances. An algorithm in SuperCollider is in charge of generating the code. Cacharpo is an example of an autonomous agent that can interact with other agents (in this case humans) with summative actions and can learn during offline training.

## Cibo and Cibo v2

Cibo (Stewart and Lawson 2019) is a VA built with interconnected neural networks that generate TidalCycles code in a solo performance style using samples from a training corpus. An encoder-decoder sequence-to-sequence architecture, typically used for language translation, is implemented using the PyTorch library. The training was based on recordings of TidalCycles performances by the authors Stewart and Lawson. Some open questions emerged from the results: *"When live-coding's intent is to show the work, what does it mean if Cibo shows its work and yet no-one, not even the creators, understand entirely how it [is] working?"* (Stewart and Lawson 2019). Cibo is said to fulfil Tanimoto's level 5 of liveness, *"tactically predictive"*. Cibo v2 (Stewart and Lawson 2020) is a VA implemented in TidalCycles that builds on Cibo. This version adds more neural networks modules to improve the performance (e.g. progression and variation). The agent has been trained with recordings of performances by several live coders using TydalCycles and operates based on characteristics learned from the training material: *"The resulting performance agent produces TidalCycles code that is highly reminiscent of the provided training material, while offering a unique, non-human interpretation of TidalCycles performances."* (Stewart and Lawson 2020, 20). Cibo and Cibo v2 are examples of autonomous agents that can learn during offline training. Although it is reported the intention to explore the agent as a co-performer, at the moment the agent performs solo.

## Autopia

Autopia (Lorway et al. 2019) is a VA that can participate in a collaborative live coding performance using the collaborative live coding tool Utopia, a SuperCollider library for communicating and sharing code. The agent generates code based on genetic algorithms, where multiple generations of agents are created. The fitness evaluation function required to evaluate the fitness of the population members is based on machine listening, where the more similar the audio output of the other performers, the higher the fitness will be. For selecting the next generation of programs, gamification elements are added to the collaborative environment in which the agent participates. A voting system is included where it is expected that humans and VAs can vote for each other. Autopia was demonstrated at the Network Music Festival 2020 using Extramuros for the network collaboration [[6. https://networkmusicfestival.org/programme/performances/electrowar-flux]]. Autopia exemplifies an autonomous agent that can interact with other agents with multiplicative actions and can learn using genetic programming.

### Mégra

The Mégra System (Reppel 2020) is a stochastic environment based on Probabilistic Finite Automata as a data model. The system is developed with Common Lisp and SuperCollider is used as a sound engine. The system is designed to work with a small data set that is trained using machine learning techniques and which allows for real-time interaction. It is possible to create sequence generators using different techniques (e.g. transition rules, training and inference, manual editing, and so on). The model can be visualized in real time as the code is updated. Mégra exemplifies an autonomous agent that can learn during online training. It also represents cooperation with the human live coder à la musical cyborg, where a continuous dialogue between the human live coder and the system creates a sense of unified identity.

## Research Directions

### Further Research Ideas

From the analyzed readings, a number of further research ideas emerge, which we outline next.

Out of all the VAs investigated here, only the Betablocker (Griffiths 2007, Bovermann and Griffiths 2014) has explicitly explored live coding beyond the laptop and keyboard using a gamepad and a Gameboy DS system. Researching this topic further seems of interest, which could be in alignment with the Internet of Things and smart objects research (Fortino and Trunfio 2014). A major criticism of performing music with personal computers and the genre of laptop music is the lack of bodily interactions as well as the lack of transparency of the performer's action. Live coding represents a step forward to avoid obscurantism by projecting the performer's screen showing the code. However, the performer's action can still be difficult to understand for an audience who is not code-literate. Emphasizing the legibility of the code is relevant so that it is clear at all times the processes and decisions taken by both the VA live coder and also the human live coder. It is still an open question how to find the right balance between simplicity and complexity: *"Live coding languages need to avoid unnecessary detail while keeping interesting possibilities open."* (Griffiths 2007, 179)

Inspired by the human-centred approach of using machine learning algorithms as a creative musical tool (Fiebrink and Caramiaux 2018), the Mégra system (Reppel 2020) brings the training process of machine learning to the live coding performance so that it is also part of the 'algorithmic thinking' of the live coder. The online machine learning training exposed in this system contrasts with the offline training process of machine learning of Cacharpo (Navarro and Ogborn 2020), Cibo (Stewart and Lawson 2019), and Cibo v2 (Stewart and Lawson 2020). In a survey about the design of future languages and environments for live coding (Kiefer and Magnusson 2019), the respondents highlighted the following features for a live coding language for machine listening and machine learning: flexibility, hackability, musicality, and instrumentation of the machine learning training. It would be interesting to compare the same system using both online and offline learning to identify any musical and computational differences. Sema is a user-friendly online system that allows practitioners to create their live coding language and explore machine learning in their live coding practice (Bernardo et al. 2020). The findings from a workshop with Sema indicated that the topics of programming language design and machine learning can be challenging for beginners in computer science. It is an open question how we can integrate machine learning concepts into the legibility of the code during a live coding session.

Another open question is how to evaluate these new algorithms from a musical and computational perspective (Xambó et al. 2019). Who should evaluate these systems, the audience, the live coder, or the virtual agent? How should these systems be evaluated? As previously discussed, there are several theoretical frameworks and approaches that we can borrow, but the literature on evaluating VAs in live coding is scarce. A point of criticism of these systems is the lack of formal evaluation methods and the need to provide better documentation for supporting reproducible research (Wilson et al. 2020). An effort should be continued to put into documenting and publishing data and code so that the community can progress faster.

Exploring a laptop ensemble with only VAs is proposed from LOLbot's research findings (Subramanian et al. 2012). As discussed earlier, we envision multiple agent collaboration or VA-VA collaboration as an interesting research space (Xambó et al. 2017). However, the results of a survey on music AI software (Knotts and Collins 2020) point that VAs should not replace humans, as human creativity is difficult to model. A promising area of research is the alternative roles that VAs can take beyond imitating humans: *"tools that take-over or control the creative process are of less interest to music creators than open ended tools with many possibilities."* (Knotts and Collins 2020, 504)

## Speculative Futures

One speculative future is facing a time when the virtual agents are completely autonomous, become independent of humans and create their own communities (Collins 2011). This raises potential machine ethics that should be considered when designing future generations of VAs. Isaac Asimov's three laws of robotics are a valuable precursor for this potential societal challenge (Asimov 1950).

When interacting with humans (HA-VA), what role do we envision for the VAs? Intelligent tutors, who can help the learner to develop musical and computational skills? Virtual musicians, who can create new unimaginable music? Co-creative partners, who are always available to rehearse and perform? Computational-led companions, who can speed up the process of live coding? Tools to understand how human live coders improvise together by modelling their behaviours? (Collins 2011, Subramanian et al. 2012, Xambó et al. 2017).

If/when we reach the *"strategically predictive"* level 6 of liveness in the process of programming (Tanimoto 2013), what will be the role of the human live coder? If the VA completes our code or even writes the code on our behalf, can this still be considered collaborative music live coding? Will this new approach to music-making change the musical aesthetic? (Kiefer and Magnusson 2019). Is the intention to create a VA that produces a live coding performance indistinguishable from a human live coder and passes the Turing test? (McLean and Wiggins 2010) Are we going to be able to produce VAs who are more responsible for the creative outputs than at present? (Wiggins and Forth 2018)

# Conclusion

In this article, we presented a short review of past, present, and future perspectives of research and practice on virtual agents in live coding. We grouped a set of selected references into two main categories: (1) *conceptual foundations;* and (2) *practical examples*. In conceptual foundations, we outlined key terms and theoretical frameworks that can be helpful to understand VAs in live coding. In practical examples, we described exemplary examples of VAs in live coding based on two dimensions, with the assumption that the VAs are autonomous: (1) *social interactivity*; and (2) *learnability*. We concluded by envisioning further research ideas and speculative futures.

Live coding is a promising space to explore machine learning and computational creativity. A range of perspectives can be researched and imagined of what reality a VA live coder represents. We are just at the beginning of the journey with more open questions than answers. It is in our hands to shape it in beneficial ways to the environment and world we live in.

## Biography

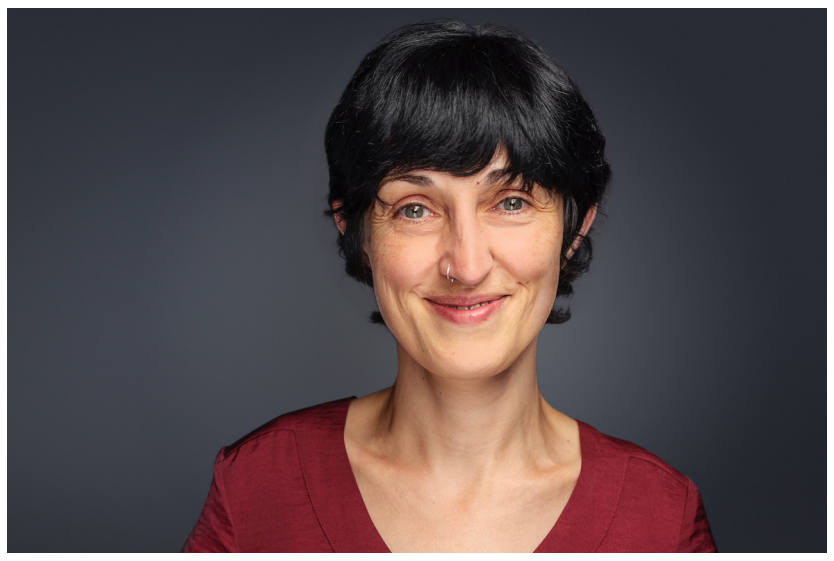

**Anna Xambó | http://annaxambo.me**

Anna Xambó is a Senior Lecturer in Music and Audio Technology at De Montfort University, a member of Music, Technology and Innovation - Institute of Sonic Creativity (MTI$^2$) and an experimental electronic music producer. Her research and practice focus on new interfaces for music performance looking at live coding; collaborative and participatory music systems; and multichannel spatialization. She is PI of the EPSRC HDI Network Plus funded project “MIRLCAuto: A Virtual Agent for Music Information Retrieval in Live Coding” (https://mirlca.dmu.ac.uk), investigating the use of a live coder virtual agent and the retrieval of large collections of sounds. Her solo and group performances have been presented internationally in China, Denmark, Germany, Norway, Spain, Sweden, UK, and USA. She has taken leading roles in organisations with a special interest in improving the representation of women in music technology: Women in Music Tech (2016-2017, Georgia Tech); WoNoMute (2018-2019, NTNU/UiO); and WiNIME (2019-present).
http://annaxambo.me